\documentclass{osa-article}
\journal{osajournal}
\articletype{Research Article}

\usepackage{lineno}
\usepackage{amsmath,amsfonts}
\usepackage{graphicx}
\usepackage{setspace}
\usepackage{tocloft}
\begin{document}

\title{Computational ghost imaging for transmission electron microscopy}

\author{Akhil Kallepalli,\authormark{1,*} Lorenzo Viani,\authormark{2,3} Daan Stellinga,\authormark{4} Enzo Rotunno,\authormark{3,**} Ming-Jie Sun,\authormark{5} Richard Bowman,\authormark{6} Paolo Rosi,\authormark{2} Stefano Frabboni,\authormark{2} Roberto Balboni,\authormark{7} Andrea Migliori,\authormark{7} Vincenzo Grillo,\authormark{3} and Miles Padgett\authormark{1}}

\address{\authormark{1}School of Physics and Astronomy, University of Glasgow, Glasgow, G12 8QQ, UK\\
\authormark{2} Department of Physics, IT and Mathematics, University of Modena and Reggio Emilia\\
\authormark{3} CNR-NANO Via G. Campi 213/a, I-41125 Modena, Italy \\
\authormark{4} Faculty of Science and Technology, University of Twente, The Netherlands\\
\authormark{5} School of Instrumentation and Optoelectronic Engineering, Beihang University, Beijing 100191, China\\
\authormark{6} Department of Physics, University of Bath, BA2 7AY, UK\\
\authormark{7} CNR-IMM Via Gobetti 101, 40129 Bologna, Italy \\}

\email{\authormark{*} akhil.kallepalli@glasgow.ac.uk} 
\email{\authormark{**} enzo.rotunno@nano.cnr.it} 

\begin{abstract}
While transmission electron microscopes (TEM) can achieve a much higher resolution than optical microscopes, they face challenges of damage to samples during the high energy processes involved. Here, we explore using computational ghost imaging techniques in electron microscopy to reduce the total required intensity. The technological lack of the equivalent high-resolution, optical spatial light modulator for electrons means that a different approach needs to be pursued. To this end, we show a beam shaping technique based on the use of a distribution of electrically charged metal needles to structure the beam, alongside a novel reconstruction method to handle the resulting highly non-orthogonal patterns. Second, we illustrate the application of this ghost imaging approach in electron microscopy. To test the full extent of the capabilities of this technique, we realised an analogous optical setup method. In both regimes, the ability to reduce the amount of total illumination intensity is evident in comparison to raster scanning. 
\end{abstract}

\section{Introduction}
Scanning transmission electron microscopes (STEM) achieve high resolution imaging by raster scanning a focused beam of electrons over the sample and measuring the transmission to form an image. Raster scanning is a simple approach to creating an image in the absence of complicated detector arrays. An alternative approach that has been pioneered in the optical regime has been measuring the transmission of the sample when illuminated with more complicated spatial patterns. This approach, often referred to as classical computational ghost imaging, inverts knowledge of the known patterns and their measured transmission to reveal the image. Furthermore, the inversion can be constrained using known image properties (e.g. sparsity of spatial frequencies) such that the image can be estimated from fewer measurements than would have been required by a raster scan; a form of compressed sensing \cite{Sen2005,Duarte2008}.

The spatial structuring of optical beams has been extended to matter waves, primarily electrons. This structuring of electron beams paved the way for the production of vortex electron beams \cite{Uchida2010,Verbeeck2010,McMorran2011}, self-accelerating beams \cite{Goutsoulas2021}, non-diffracting beams \cite{Voloch-Bloch2013,Grillo2014, Grillo2016} and orbital angular momentum analysers \cite{Grillo2017, Tavabi2021}. In the case of electron beams, an arbitrary way of flexibly spatially programming a beam is problematic and existing methods for the dynamic shaping of electron beams are limited. However, recent progress could inspire to build a tuneable phase plate based on a set of needle-shaped electrodes producing a continuous phase landscape. In this work, we use electron beam shaping technologies to demonstrate that the technique can be applied in electron microscopy and show it is possible to create complicated far-field patterns that when combined with a novel inversion algorithm produce images comparable to those produced by raster scanning but using much lower peak intensities. The technique is firstly developed with a light-optics analogue toy experiment and then deployed in the electron microscope real measurement.

The potential advantages of the computational ghost imaging approach are two-fold. First, that single element detectors usually have better characteristics, e.g. for light an extended range of operating wavelengths, for electrons a larger quantum efficiency. Second, the flexibility of the illuminating patterns make them potentially more suitable for imaging optimisation by compressed sensing algorithms, reducing the number of required measurements \cite{Wakin2006,Candes2008,Romberg2008}. This ability to image with fewer measurements potentially increases the frame rate and/or reduces the exposure of potentially delicate samples to the illumination. Using these computational ghost imaging techniques, novel imaging systems have been developed that include imaging at terahertz wavelengths \cite{Stantchev2016}, time of flight imaging \cite{Sun2019,Howland2011}, and imaging sensitive to fluorescent lifetime \cite{Futia2011}. 

In addition to optical wavelengths, electron beams can also be used for precision imaging. Since the time-independent wave equation for the light and electrons are perfectly analogous \cite{Shiloh2014}, components like lenses for electron beams are defined with a similar nomenclature as for light. However, optical elements for light are inherently easier to build and spatial light modulators (SLM) that shape the optical beams in a programmable, and near arbitrary way, do not explicitly exist for electrons. Although single-element detector-based imaging has previously been introduced for specialised electron imaging systems \cite{Shwartz2021}, the barrier to more general use is absence of SLM-like technologies.

In the context of electron beams, holographic plates \cite{Shiloh2014, Rosi2022} laid the foundation for applications related to the shaping of electron beams. However, these are passive elements and cannot be dynamically addressed like optical SLM. A most welcomed recent development has been the introduction of active optical elements or programmable phase plates \cite{Verbeeck2018}. These are based on simple electrostatic elements to be positioned along the beam path. Two technological solutions are currently available: the research group led by Verbeeck has been developing devices based on Einzel lenses. These lenses work like the individual pixels of an optical spatial light modulator but the number of addressable points are few in comparison the the optical devices \cite{Verbeeck2018}. An additional challenge for these lenses arises from the difficulty to bias and drive many connections from outside the microscope column, which necessarily limits the useful fill fraction of such a pixel-based approach. As an alternative, other groups are using separated electrodes to define specific phase landscapes across a wider area. This electrode-based approach has the benefit that the beam itself only interacts with the electric field created rather than the modulator components themselves, avoiding the main problems of the other approach. However, whereas the solution is defined everywhere the phases $\phi$ are limited by the harmonic condition $\nabla ^2 \phi=0$ everywhere except on the electrodes \cite{Ruffato2020}. In other words, the variations in possible phase modulation is far more restricted.

The most trivial electrode used thus far is a simple metallic needle \cite{Griffiths1996}. The definition of the electrostatic field around a charged microtip has been studied in detail \cite{Matteucci1992,Beleggia2014}, wherein charge applied to these needles induces a phase shift around their tips to the electron beam. Interestingly, the phase shift generated by a metallic needle was used as key to implement an orbital angular momentum (OAM) sorter \cite{McMorran_2017}. More importantly, multiple metallic needles have been demonstrate to be capable of producing complex tunable caustic phenomena in the far field\cite{Tavabi2015}. This current work uses an arrangement of six such needles.% as shown in figure \ref{fig:6_pin_Images}.

\section{Computational ghost imaging in TEM}
\begin{figure}[ht]
\centering
\includegraphics[width=0.95\columnwidth]{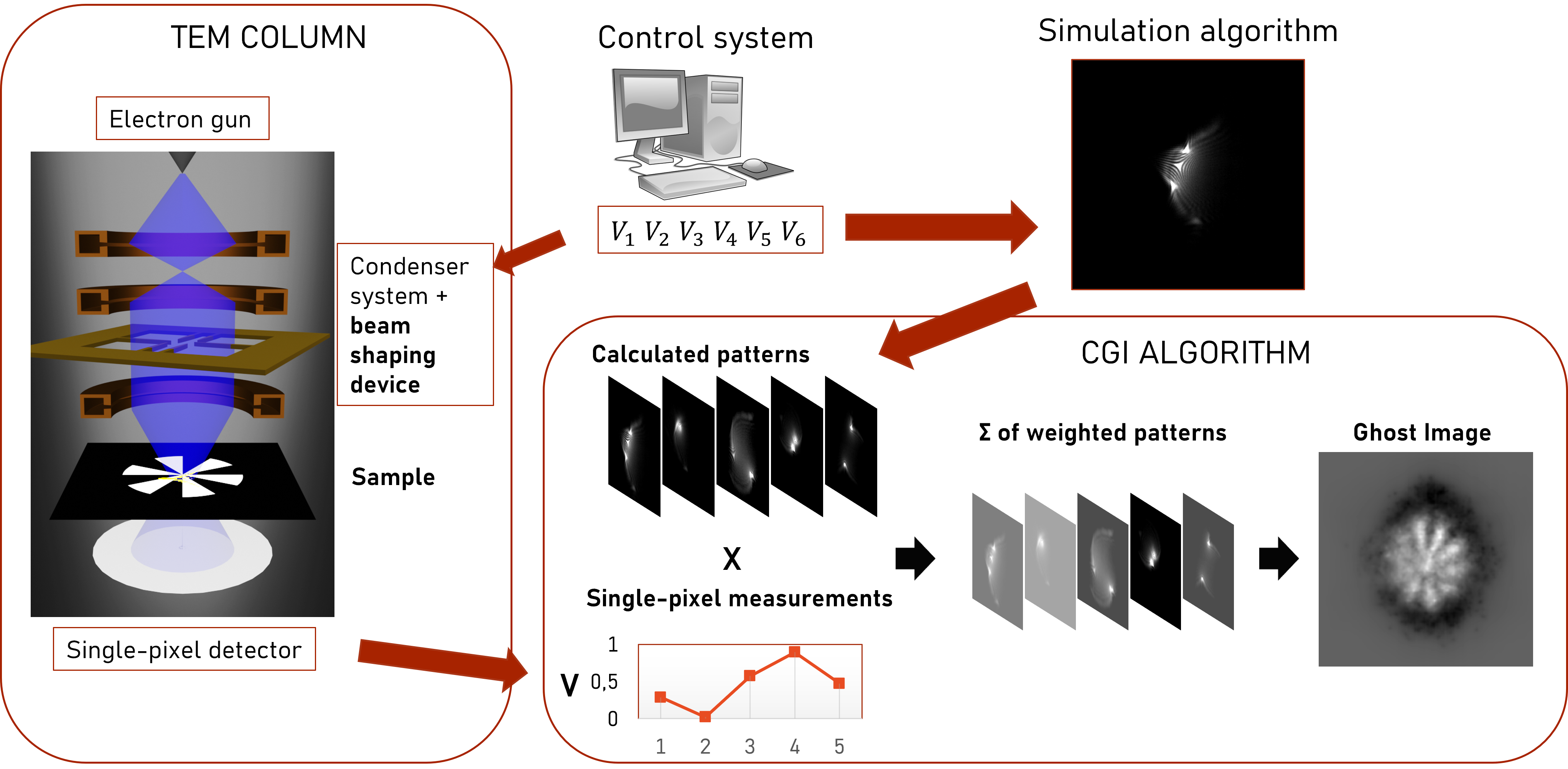}
\caption{A TEM sample holder carrying our needle-based beam-shaping device is inserted in the column at the condenser system level. Dedicated electronics are used to randomly select six random biases and apply them to the needles. The electron beam crosses the potential distribution and is imaged on the sample. Then, the total transmitted intensity is acquired by a single-pixel detector, whose output value is acquired by the electronics and recorded together with the six biases. After that, the patterns used for the measurement are calculated starting from the six biases. Finally, the combination of the single-pixel measurements and the calculated patterns through the CGI algorithm yields the ghost image of the sample.}
\label{fig:Electron_setup}
\end{figure}

We propose a design of the experimental setup for TEM implementation of Computational Ghost Imaging. Figure \ref{fig:Electron_setup} shows a schematic representing the key steps of the experiment. 

\begin{figure}[ht]
\centering
\includegraphics[width=0.95\columnwidth]{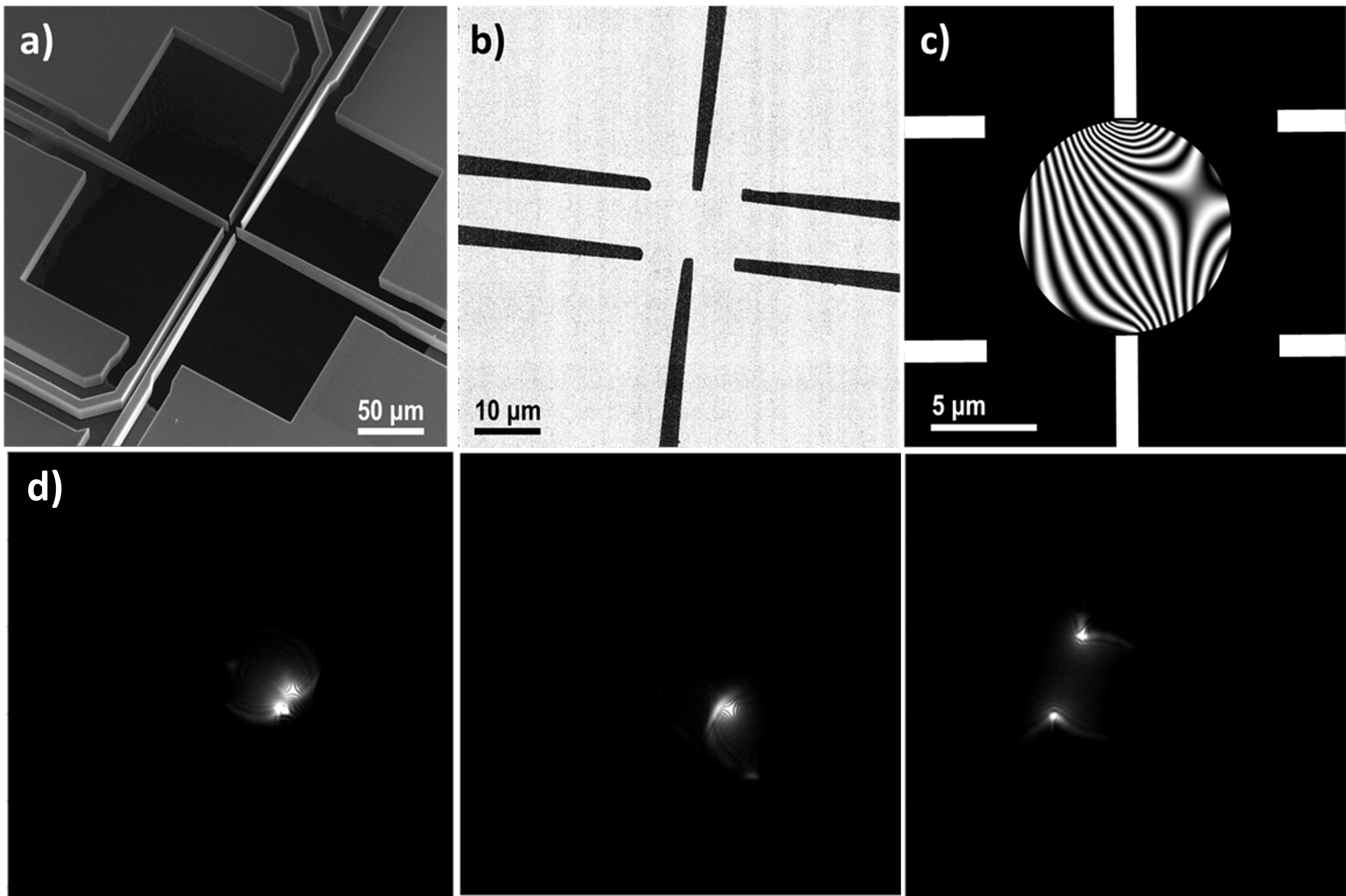}
\caption{SEM images (a) and (b) illustrate the six-needle device and its active region. The active region is limited by a 10 $\mu$m diameter aperture and a resulting simulation of a typical phase profile produced by these needles is shown in (c). The white bars in (c) show the positions of the needles. (d) shows three TEM experimental images of the intensity distribution in the far-field produced by the device.}
\label{fig:6_pin_Images}
\end{figure}

In complete analogy with the optical realisation of ghost imaging, the technique requires a way to reliably produce a form of structured illumination. As mentioned before, in optical systems this is possible thanks to SLM technology, but there is not any equivalent form of beam-shaping technology available for electrons. Thus, we developed a device that uses six electrically biased needles to generate a potential distribution, as shown in figure \ref{fig:6_pin_Images}. When the device is inserted into the TEM column the potential distribution interacts with the electron beam and applies a position-dependent phase shift to it (see figure \ref{fig:6_pin_Images}(c)). 

The beam shaping device produces complex patterns in far field observation, with a shape dependent on the six biases applied to the needles. Thus, our device allows us to control the illumination, albeit with far fewer degrees of freedom than those available with a SLM. Some examples are shown in figure \ref{fig:6_pin_Images}(d). Notice how there can be a single intensity peak or more than one with variable separation, from superimposed to very far apart.

A critical limitation of the patterns produced by the beam-shaping device is that they are highly non-orthogonal. In stark contrast, SLMs can produce completely orthogonal pattern sets, like the Hadamard basis, which are inherently better for CGI. To this end, we also developed a modified CGI algorithm that adds orthogonalisation to the image reconstruction process. %The details will be described in the next section.

%Using our newly built toolset, we prepared a optical mockup experiment to check the performance of the modified CGI algorithm, then we made a first attempt to perform the experiment in a TEM. Figure \ref{fig:Electron_setup} shows a simple schematic of the setup used for TEM CGI and the key steps of the technique. The next sections will provide a description of the modified CGI algorithm, its performance against traditional CGI measured with an optical toy experiment that reproduces the TEM implementation and finally the first TEM ghost image that we managed to reconstruct.

\section{Novel algorithm for ghost imaging with non-orthogonal pattern basis}
The choice of mask patterns is key to the performance of the system as it defines the resolving capability of the system and the applicability of the reconstruction algorithm. The field of ghost imaging has seen an evolution in both experimental setups and algorithms \cite{Ferri2010,Gibson2020}. Although the vast majority of computational ghost imaging and the related work on single pixel cameras use a spatial modulator positioned in the image plane of the object, the original work by Shapiro \cite{Shapiro2008} recognised that this need not be the case. Rather, given any specific modulation, the resulting form of the light field in the plane of the object could be calculated using numerical beam propagation techniques enabling both lens free and far field implementations \cite{Kallepalli2021_SciReps}. 

Another important aspect of ghost imaging are the algorithms used to recover an image from the measurements and knowledge of the patterns used. This problem can be cast into a simple matrix form:
\begin{equation} 
    A\vec{x}=\vec{b},
    \label{eq:linSys}
\end{equation}
where $A$ is a matrix containing the patterns $\vec{p_{i}}^\top$ as rows, $\vec{x}$ represents the (discretised) scene, and $\vec{b}$ is the set of measurements $b_{i}$. The methods used to retrieve $\vec{x}$ and thereby an image fall roughly into two categories, offline or online \cite{Karp1992}. Offline approaches infer the image from the final complete set of patterns and corresponding measurements. The majority of well known methods for solving linear systems fall in this category, e.g. the matrix (pseudo)inverse, Newton's method, etc. Conversely, online approaches update a best guess of the image with every new measurement and light pattern. The most common online algorithm involves a straightforward weighted sum of the patterns, where the weights are adjusted based on a statistical interpretation of the measurements:
\begin{equation}
        \vec{g}_{i} = \vec{g}_{i-1} + (b_{i} - \langle b \rangle _{i}) \vec{p}_{i}.
    \label{eq:TGI}
\end{equation}
Here $\vec{g}_{i}$ is the best estimate after $i$ patterns and $\langle b \rangle _{i}$ indicates the mean detector value up to the current pattern, and $\vec{g}_{0}=\vec{0}$. This algorithm, which we will refer to as traditional ghost imaging (TGI), has been extensively explored with different setups \cite{Gatti2004,Sun2012}.

The primary advantage of online reconstruction algorithms lies in the low computational overhead and robustness to under-sampling. In the absence of additional constraints, to achieve a reasonable spatial resolution, the number of illumination patterns required is typically equal to the number of pixels in the image. This quickly makes the matrix $A$ from equation \ref{eq:linSys} very large, leading to high memory requirements and long computation times for offline methods. Online methods only need to store the current best guess of $\vec{x}$ and the current pattern, can run in parallel with the measurement, and only ever operate on vectors of limited size. Moreover, TGI is by its very nature impervious to problems with under-sampling and can show approximations to the correct image with comparatively only a handful of measurements. 

Both classes of reconstruction method work significantly better when used with orthogonal pattern sets, for example Hadamard matrices \cite{Pratt1969,Sloane1976,Zhang2017}, where each mask probes a subset of spatial frequencies. For offline methods this is equivalent to ensuring full row rank, which increases the stability of the matrix inversion results. For TGI, it ensures that each pattern's spatial frequencies are counted democratically in the mean correction term. Either way, in the absence of noise, the result approaches the true image when the number of orthogonal patterns equals the number of pixels.

\begin{figure}[t]
\centering
\includegraphics[width=0.9\columnwidth]{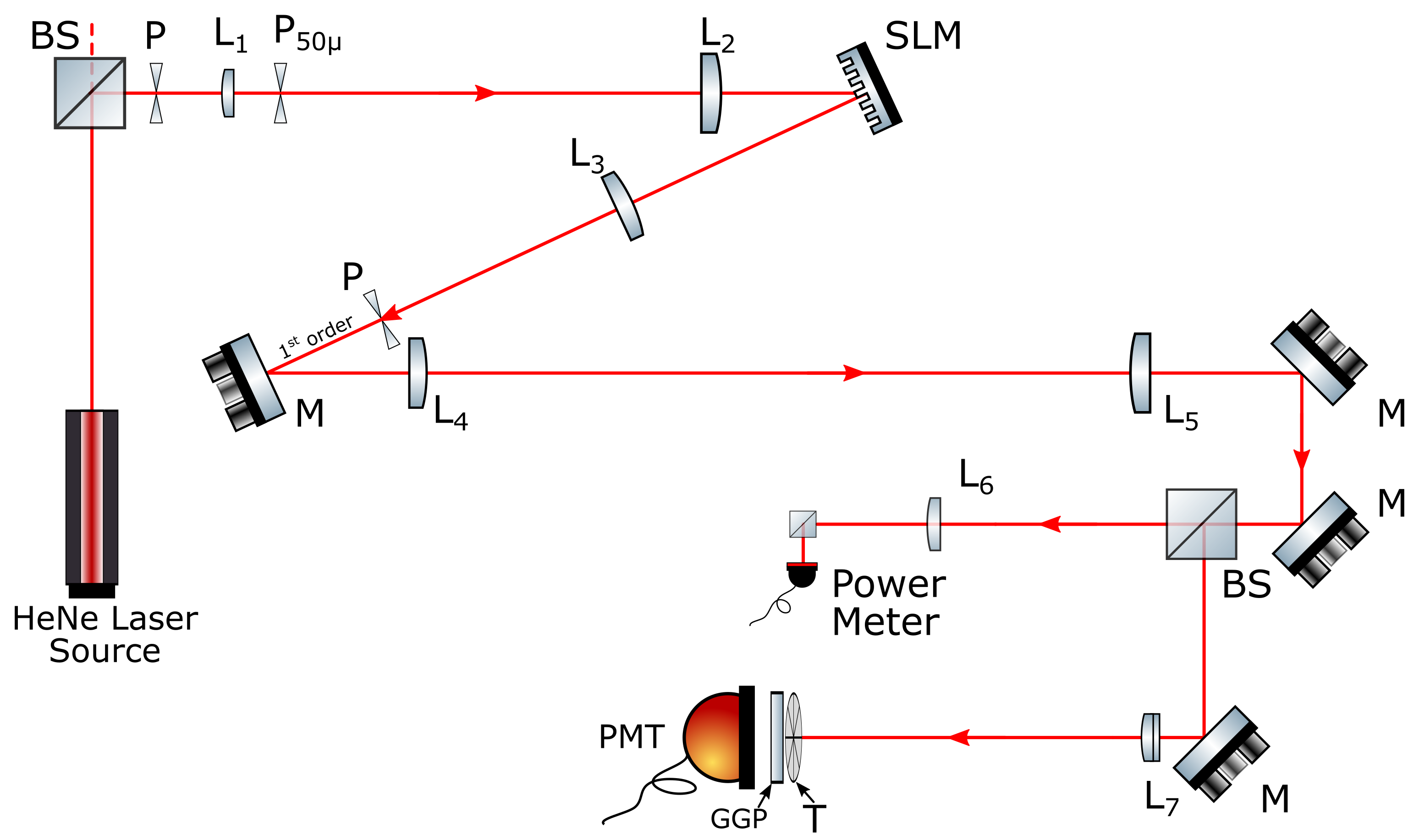}
\caption{Light from a HeNe at 633 nm is polarised by a polarising beam splitter (PBS), then spatially filtered and expanded to fill the aperture of an SLM using an aperture (P), a 50 mm focal length lens (L$_1$), precision pinhole (P$_{50 \mu}$) and 400 mm focal length lens (L$_2$). The SLM displays a superposition of the six-needles phase masks and a carrier diffraction grating. The SLM is optimised to project the majority of the modulated light into the first diffraction order, which is selected using a 250 mm focal length lens (L$_3$) and aperture (P) placed in the Fourier plane of the SLM. This light is propagated into the far-field through a combination of lenses (250 mm focal length lens (L$_4$) and 400 mm focal length lens (L$_5$)). The beam is propagated into two separate arms using beamsplitter (BS). One arm uses a power meter to measure the laser power and the second arm consists of the target (T). The photomultiplier tube (Thorlabs, PMM02) measures the transmitted light after interaction with the target. Author Figure adopted from Kallepalli \textit{et al.} \cite{Kallepalli2021Ghost}}
\label{fig:ExperimentalDesign}
\end{figure} 

If the mask patterns are implemented using a pixelated spatial light modulator, specifying their design to be orthogonal is straightforward (e.g. Hadamard, or similar). If, however, the patterns are produced in other ways, e.g. using natural scattering or with the highly limited modulators available in a TEM, then orthogonality between the patterns can no longer be enforced and the standard reconstruction algorithms no longer converge to an accurate image. The TGI algorithm in particular will overemphasise the overlapping parts of the patterns. To resolve this issue, we report an alternative approach that is computationally efficient and makes optimal use of patterns regardless of their orthogonality. This alternative method, which we will call orthogonalised ghost imaging (OGI), takes inspiration from the Kaczmarz algorithm for solving linear systems \cite{Popa2004,Bautu2006,Strohmer2008,Pires2016,Sun2020}.

OGI essentially takes the projection of each new pattern onto the current best pattern estimate as an extra correction to the mean detector value used in TGI:
\begin{equation}
        \vec{g}_{i} = \vec{g}_{i-1} + \left[\left(b_{i}-\langle b\rangle_{i}\right)-\left(c_{i}-\langle c\rangle_{i}\right)\right] \vec{p}_{i},
    \label{eq:OTGI}
\end{equation}
where $c_{i}=\vec{p}_{i}\cdot\vec{g}_{i-1}$ is the predicted signal under the assumption that the previous best estimate is the correct reconstruction. This has the effect of mitigating double counting stemming from non-orthogonality and emphasising actual new information present in the measurement. Note that care needs to be taken such that both the measured and predicted values are correctly scaled to each other such that the reconstructed image approaches the true image. This is equivalent to ensuring that the quantity $[(b_{i}-\langle b \rangle_{i})-(c_{i} - \langle c \rangle_{i})]$ tends to zero. The approach we adopt here to ensure this scaling is to iteratively normalise the signals such that $\langle c \rangle = \langle b \rangle$ and $\sigma_{c} = \sigma_{b}$. 

%Both algorithms, by principle, deal with noise differently. The TGI algorithm's averaging approach decreases the noise with larger pattern sets. If the illumination budget is unlimited, the noise decreases with increased sampling (as a function of $1/\sqrt{N}$ when $N$ denotes the number of measurements). In contrast, once the OGI algorithm has converged on the image, further measurement will alter the noise but in general will not reduce it. OGI is therefore particularly appropriate for use when the illumination budget is limited. If longer measurements are possible, a relaxation term could be introduced into OGI to slow down convergence and bring its sensitivity to noise in line with TGI, without losing the ability to work with non-orthogonal patterns. 

\section{Optical ghost imaging}
The modulation device investigated in this article is inspired from the charged microtips approach, designed by Matteucci \textit{et al.} \cite{Matteucci1992}. The needles, to which random voltages are applied, create a spatially structured phase change in the transmitted electron beam, leading to a complicated, spatially structured, point spread function in the far-field that illuminates the sample. An example is shown in figure \ref{fig:6_pin_Images}. In our experiment, the equivalent phase change to that induced by the charged needles is, in the optical regime, created using a liquid crystal spatial light modulator (LC-SLM). 

Our analogous optical experimental set up \cite{Kallepalli2021Ghost} for demonstrating the six-needles approach to computational ghost imaging is shown in Figure \ref{fig:ExperimentalDesign}. The expanded output beam from a HeNe laser is shaped using randomly switched six-needles that generate phase masks. These phase masks are displayed on the spatial light modulator (SLM), structuring the laser beam and propagating complex intensity patterns into the far-field.

Reflective SLMs have the advantage of faster response rates and a higher fill factor than those devices based upon transmission. However, these reflective devices are susceptible to manufacturing limitations that result in deviations from flatness of the addressed surface, resulting in aberrations such as astigmatism and defocus. Correction of these distortions is key to accurately manipulating the phase of the incoming light beam. \cite{Jesacher2007,Cizmar2010}. Our approach to aberration correction, adapted from Turtaev \textit{et al.} \cite{Turtaev2017}, uses a diagonal blazed phase grating with a predetermined periodicity. The aberration-corrected SLM operates in a diffractive mode, and therefore, the masks are combined with a phase-grating to diffract the desired pattern to the first-order which is selected by a pinhole in the Fourier-plane. The chosen order is then propagated through the lens system and into two optical paths (Figure \ref{fig:ExperimentalDesign}): (a) laser power meter to monitor the laser power for any variations and (b) ghost imaging system (photomultiplier tube, PMT). The PMT is positioned to collect the light after transmission by the object. Based on the known voltage applied to each of the needles and the phase change this creates, the calculation of the far-field intensity distribution and the measured signal are used to reconstruct the image, using the various algorithms described above.

\begin{figure}[t]
\centering
\includegraphics[width=0.95\columnwidth]{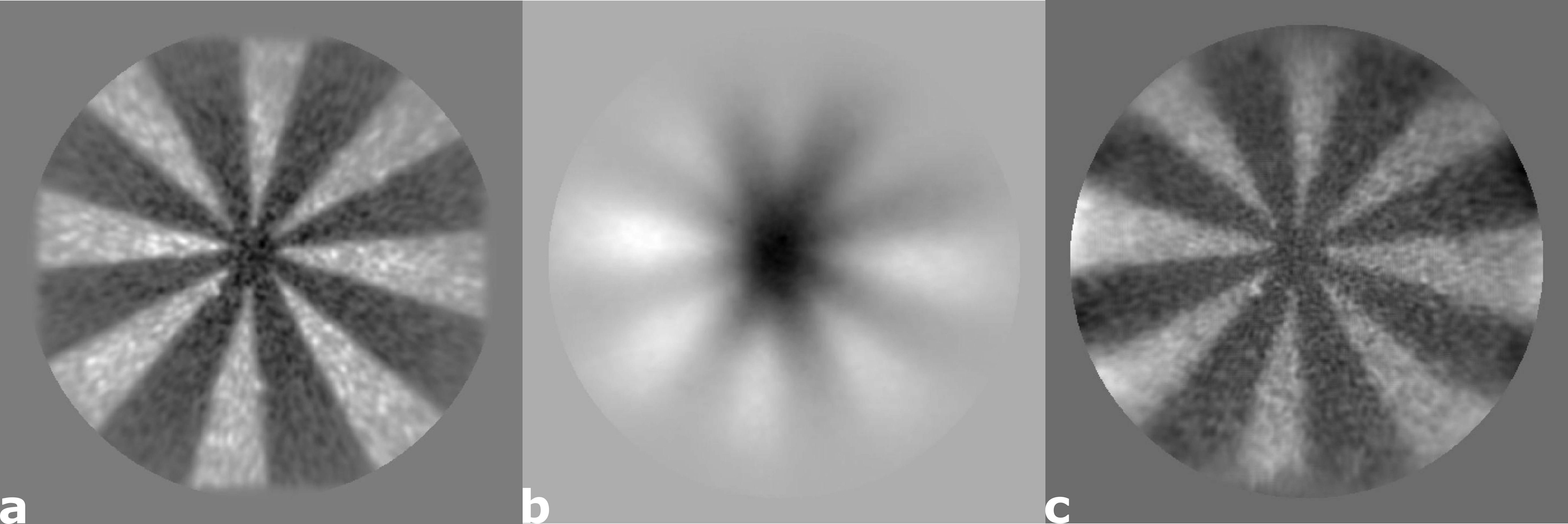}
\caption{Our approach illustrates the viability of using TEM-inspired phase masks for image reconstruction using a new approach to ghost imaging known as orthogonalised ghost imaging (OGI). This approach is ideal for cases such as six-needles phase mask sets, where the pattern set is not orthogonal. The comparison shows (a) randomised raster scan using traditional ghost imaging (TGI), (b) six-needle phase masks using TGI and (c) six-needle phase masks using the proposed OGI technique. In each case, the construction was done using a set of 50000 patterns and the reconstructed images are of 512 $\times$ 512 pixel resolution. The binary amplitude target used is a laser cut 8 spoke shape.}
\label{fig:Comparison}
\end{figure}

\begin{figure}[h]
\centering
\includegraphics[width=0.8\columnwidth]{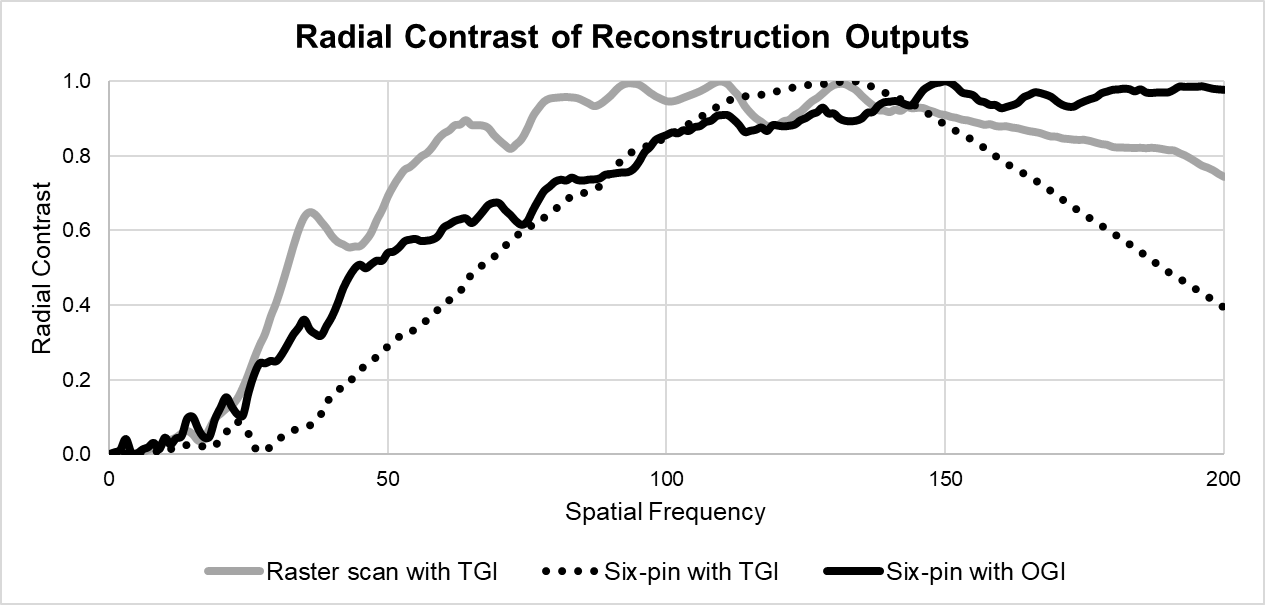}
\caption{Radial contrast functions associated with the images shown in figure \ref{fig:Comparison}, quantifying the resulting reconstructions using OGI-six needles, TGI-six needles and TGI-raster scans, each using 50000 patterns. The six-needles patterns are non-orthogonal by nature.}
\label{fig:Radial_Contrast}
\end{figure}

We benchmark the performance of our six-needle approach against a traditional raster scan. Raster scanning is achieved by displaying the appropriate series of linear phase gratings on the SLM to produce a scanning spot in the far-field. The experiment results are presented in figure \ref{fig:Comparison}, which gives a visual representation of the TGI and OGI algorithms performance compared to that of a raster scan. Clearly, the non-orthogonal nature of the six-needle pattern sets has significantly degraded the reconstructed image when using the TGI algorithm. By comparison we see that the OGI algorithm allows reconstruction an image with the same resolution as a traditional raster scan. As a more quantitative comparison, the spatial frequency and resolution is assessed using a radial contrast function (Figure \ref{fig:Radial_Contrast}). With this number of measurements TGI has essentially converged to its final resolution, with any further improvements primarily consisting of suppressing the noise. OGI on the other hand has not yet converged, but already significantly outperforms TGI for the non-orthogonal pattern set created by the six-needle modulator. 

Although the overall optical energy used in the six-needle OGI and the traditional raster scan is the same we note that the peak intensity in six-needle approach is orders of magnitude lower. This reduction would be extremely important in situations where the sample can be damaged by the illumination, especially if that damage mechanism was non-linear with respect to intensity. 

\section{TEM computational ghost imaging}
Having proven the applicability of computational ghost imaging with the optical analogy, we attempted an experimental demonstration with electrons. The experiments were carried out by mounting the modulation device in a biased I-V TEM sample holder. 
The sample to be imaged is a Siemens star target with 5 spokes, located in the Selected Area plane of the microscope. It was fabricated on a commercially available 200 $nm$ thick Si$_3$N$_4$ membrane, covered with $\sim$150 $nm$ of gold, so that it would act as an amplitude mask. The total transmitted intensity is collected by the annular dark-field (ADF) detector, set off-axis to acquire the whole transmission. The output value of the detector is read by the control software and synchronised with the biases from the modulation device. 

% However, in preliminary measurements, the system clearly showed some form of noise that added a 5\% uncertainty on the transmitted intensity values. 
%Before the experiments, we also verified the shape of the illumination patterns with a standard CCD camera for different values of the bias. Through this approach, we proved that we can correctly predict the patterns sent to the target through a Finite Elements modelling or by training an appropriate Artificial Neural Network (ANN). 

\begin{figure}[ht]
\centering
\includegraphics[width=\columnwidth]{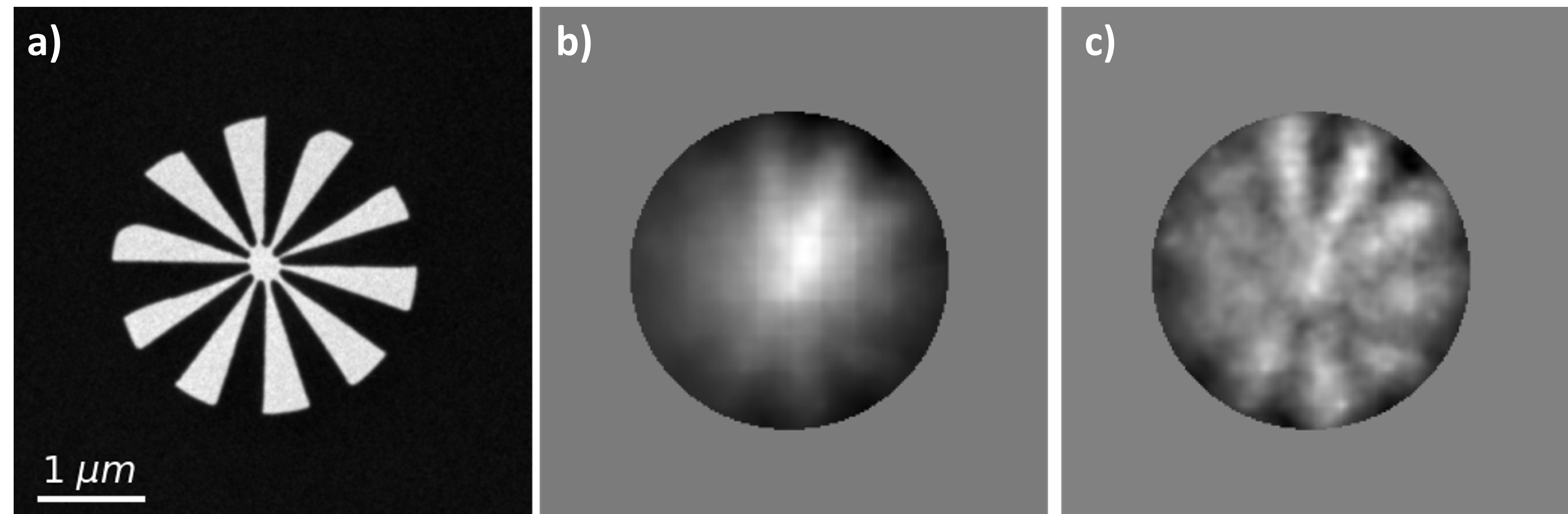}
\caption{Comparison of (a) a TEM image of the target and the images reconstructed with TGI (b) and OGI (c) algorithms. The target is a 3 $\mu$m 10-spoke Siemens star, acquired on a Gatan MSC794 CCD camera.}
\label{fig:TEM_result}
\end{figure} 

The acquisition comprised 11500 different illumination patterns. The patterns produced by the beam shaping device were not directly recorded, but instead they were recovered using a simulation algorithm after the measurement. The simulation algorithm  predicts the illumination patterns used during the experiment starting from the six biases applied to the needles of the beam-shaping device. Figure \ref{fig:TEM_result}(a) displays the target and the ghost images obtained with both the traditional (b) and orthogonalised (c) algorithms, starting from the same data. TGI is clearly outperformed by our OGI formulation (Eq. \ref{eq:OTGI}), as pointed out by the differences between the two reconstructions. The OGI image exhibits higher resolution and better contrast, resulting in an overall sharper image.

\section{Conclusions}
We have shown a successful proof-of-principle ghost imaging scheme tested in both optical and electron microscopy regimes. In the optical regime, we combine non-orthogonal intensity distribution patterns with a novel ghost imaging algorithm for image reconstruction. Due to the non-orthogonal nature of the far-field intensity pattern produced by the six-needle mask, the reconstruction requires a modified algorithm if image resolution is to be maintained. The results obtained using six-needle mask in the far-field show a comparable reconstruction resolution between the proposed method and a traditional raster scanning with a focused spot. This approach was transitioned to transmission electron microscopy, wherein the modulation method approach was used to reconstruct the object, with preliminary results showing promise for image reconstruction with reduced overall illumination intensity. 

\begin{backmatter}
\bmsection{Funding}
We wish to acknowledge the support from the European Union’s Horizon 2020 Research and Innovation Programme (Grant Agreement No. 766970 Project "Q-Sort", No. 964591 "SMART-electron"), the Royal Society, EPSRC Research Council funding to QuantIC [EP/M01326X/1] and National Natural Science Foundation of China (Grant No. 61922011 and U21B2034).

\end{backmatter}

\bibliography{sample}

\begin{thebibliography}{10}
\newcommand{\enquote}[1]{``#1''}

\bibitem{Sen2005}
P.~Sen, B.~Chen, G.~Garg, S.~R. Marschner, M.~Horowitz, M.~Levoy, and H.~P.~A.
  Lensch, \enquote{Dual photography,} {\protect\JournalTitle{ACM Trans.
  Graph.}} \textbf{24}, 745–755 (2005).

\bibitem{Duarte2008}
M.~F. Duarte, M.~A. Davenport, D.~Takhar, J.~N. Laska, T.~Sun, K.~F. Kelly, and
  R.~G. Baraniuk, \enquote{{Single-pixel imaging via compressive sampling},}
  {\protect\JournalTitle{IEEE Signal Processing Magazine}} \textbf{25}, 83--91
  (2008).

\bibitem{Uchida2010}
M.~Uchida and A.~Tonomura, \enquote{{Generation of electron beams carrying
  orbital angular momentum},} {\protect\JournalTitle{Nature}} \textbf{464},
  737--739 (2010).

\bibitem{Verbeeck2010}
J.~Verbeeck, H.~Tian, and P.~Schattschneider, \enquote{{Production and
  application of electron vortex beams},} {\protect\JournalTitle{Nature}}
  \textbf{467}, 301--304 (2010).

\bibitem{McMorran2011}
B.~J. McMorran, A.~Agrawal, I.~M. Anderson, A.~A. Herzing, H.~J. Lezec, J.~J.
  McClelland, and J.~Unguris, \enquote{{Electron Vortex Beams with High Quanta
  of Orbital Angular Momentum},} {\protect\JournalTitle{Science}} \textbf{331},
  192--195 (2011).

\bibitem{Goutsoulas2021}
M.~Goutsoulas and N.~K. Efremidis, \enquote{Dynamics of self-accelerating
  electron beams in a homogeneous magnetic field,} {\protect\JournalTitle{Phys.
  Rev. A}} \textbf{103}, 013519 (2021).

\bibitem{Voloch-Bloch2013}
N.~Voloch-Bloch, Y.~Lereah, Y.~Lilach, A.~Gover, and A.~Arie,
  \enquote{{Generation of electron Airy beams},}
  {\protect\JournalTitle{Nature}} \textbf{494}, 331--335 (2013).

\bibitem{Grillo2014}
V.~Grillo, E.~Karimi, G.~C. Gazzadi, S.~Frabboni, M.~R. Dennis, and R.~W. Boyd,
  \enquote{Generation of nondiffracting electron bessel beams,}
  {\protect\JournalTitle{Phys. Rev. X}} \textbf{4}, 011013 (2014).

\bibitem{Grillo2016}
V.~Grillo, J.~Harris, G.~C. Gazzadi, R.~Balboni, E.~Mafakheri, M.~R. Dennis,
  S.~Frabboni, R.~W. Boyd, and E.~Karimi, \enquote{{Generation and application
  of bessel beams in electron microscopy},}
  {\protect\JournalTitle{Ultramicroscopy}} \textbf{166}, 48--60 (2016).

\bibitem{Grillo2017}
V.~Grillo, A.~H. Tavabi, F.~Venturi, H.~Larocque, R.~Balboni, G.~C. Gazzadi,
  S.~Frabboni, P.-H. Lu, E.~Mafakheri, F.~Bouchard, R.~E. Dunin-Borkowski,
  R.~W. Boyd, M.~P.~J. Lavery, M.~J. Padgett, and E.~Karimi,
  \enquote{{Measuring the orbital angular momentum spectrum of an electron
  beam},} {\protect\JournalTitle{Nature Communications}} \textbf{8}, 1--6
  (2017).

\bibitem{Tavabi2021}
A.~H. Tavabi, P.~Rosi, E.~Rotunno, A.~Roncaglia, L.~Belsito, S.~Frabboni,
  G.~Pozzi, G.~C. Gazzadi, P.-H. Lu, R.~Nijland, M.~Ghosh, P.~Tiemeijer,
  E.~Karimi, R.~E. Dunin-Borkowski, and V.~Grillo, \enquote{{Experimental
  Demonstration of an Electrostatic Orbital Angular Momentum Sorter for
  Electron Beams},} {\protect\JournalTitle{Physical Review Letters}}
  \textbf{126}, 094802 (2021).

\bibitem{Wakin2006}
M.~B. Wakin, J.~N. Laska, M.~F. Duarte, D.~Baron, S.~Sarvotham, D.~Takhar,
  K.~F. Kelly, and R.~G. Baraniuk, \enquote{An architecture for compressive
  imaging,} in \emph{2006 International Conference on Image Processing,}
  (2006), pp. 1273--1276.

\bibitem{Candes2008}
E.~J. Candes and M.~B. Wakin, \enquote{An introduction to compressive
  sampling,} {\protect\JournalTitle{IEEE Signal Processing Magazine}}
  \textbf{25}, 21--30 (2008).

\bibitem{Romberg2008}
J.~Romberg, \enquote{Imaging via compressive sampling,}
  {\protect\JournalTitle{IEEE Signal Processing Magazine}} \textbf{25}, 14--20
  (2008).

\bibitem{Stantchev2016}
R.~I. Stantchev, B.~Sun, S.~M. Hornett, P.~A. Hobson, G.~M. Gibson, M.~J.
  Padgett, and E.~Hendry, \enquote{{Noninvasive, near-field terahertz imaging
  of hidden objects using a single-pixel detector},}
  {\protect\JournalTitle{Science Advances}} \textbf{2}, 1--6 (2016).

\bibitem{Sun2019}
M.-J. Sun and J.-M. Zhang, \enquote{Single-pixel imaging and its application in
  three-dimensional reconstruction: A brief review,}
  {\protect\JournalTitle{Sensors}} \textbf{19} (2019).

\bibitem{Howland2011}
G.~A. Howland, P.~B. Dixon, and J.~C. Howell, \enquote{{Photon-counting
  compressive sensing laser radar for 3D imaging},}
  {\protect\JournalTitle{Applied Optics}} \textbf{50}, 5917--5920 (2011).

\bibitem{Futia2011}
G.~Futia, P.~Schlup, D.~G. Winters, and R.~A. Bartels,
  \enquote{{Spatially-chirped modulation imaging of absorbtion and fluorescent
  objects on single-element optical detector},} {\protect\JournalTitle{Optics
  Express}} \textbf{19}, 1626--1640 (2011).

\bibitem{Shiloh2014}
R.~Shiloh, Y.~Lereah, Y.~Lilach, and A.~Arie, \enquote{Sculpturing the electron
  wave function using nanoscale phase masks,}
  {\protect\JournalTitle{Ultramicroscopy}} \textbf{144}, 26--31 (2014).

\bibitem{Shwartz2021}
S.~Shwartz, \enquote{Single-pixel imaging with high-energy electromagnetic
  radiation and particles,} {\protect\JournalTitle{Science Bulletin}}
  \textbf{66}, 857--859 (2021).

\bibitem{Rosi2022}
P.~Rosi, F.~Venturi, G.~Medici, C.~Menozzi, G.~C. Gazzadi, E.~Rotunno,
  S.~Frabboni, R.~Balboni, M.~Rezaee, A.~H. Tavabi, R.~E. Dunin-Borkowski,
  E.~Karimi, and V.~Grillo, \enquote{{Theoretical and practical aspects of the
  design and production of synthetic holograms for transmission electron
  microscopy},} {\protect\JournalTitle{Journal of Applied Physics}}
  \textbf{131}, 031101 (2022).

\bibitem{Verbeeck2018}
J.~Verbeeck, A.~Béché, K.~Müller-Caspary, G.~Guzzinati, M.~A. Luong, and
  M.~{Den Hertog}, \enquote{Demonstration of a 2x2 programmable phase plate for
  electrons,} {\protect\JournalTitle{Ultramicroscopy}} \textbf{190}, 58--65
  (2018).

\bibitem{Ruffato2020}
G.~Ruffato, E.~Rotunno, L.~Giberti, and V.~Grillo, \enquote{Arbitrary conformal
  transformations of wave functions,} {\protect\JournalTitle{Phys. Rev.
  Applied}} \textbf{15}, 054028 (2021).

\bibitem{Griffiths1996}
D.~J. Griffiths and Y.~Li, \enquote{Charge density on a conducting needle,}
  {\protect\JournalTitle{American Journal of Physics}} \textbf{64}, 706--714
  (1996).

\bibitem{Matteucci1992}
G.~Matteucci, G.~Missiroli, M.~Muccini, and G.~Pozzi, \enquote{Electron
  holography in the study of the electrostatic fields: the case of charged
  microtips,} {\protect\JournalTitle{Ultramicroscopy}} \textbf{45}, 77--83
  (1992).

\bibitem{Beleggia2014}
M.~Beleggia, T.~Kasama, D.~J. Larson, T.~F. Kelly, R.~E. Dunin-Borkowski, and
  G.~Pozzi, \enquote{Towards quantitative off-axis electron holographic mapping
  of the electric field around the tip of a sharp biased metallic needle,}
  {\protect\JournalTitle{Journal of Applied Physics}} \textbf{116}, 024305
  (2014).

\bibitem{McMorran_2017}
B.~J. McMorran, T.~R. Harvey, and M.~P.~J. Lavery, \enquote{Efficient sorting
  of free electron orbital angular momentum,} {\protect\JournalTitle{New
  Journal of Physics}} \textbf{19}, 023053 (2017).

\bibitem{Tavabi2015}
A.~H. Tavabi, V.~Migunov, C.~Dwyer, R.~E. Dunin-Borkowski, and G.~Pozzi,
  \enquote{Tunable caustic phenomena in electron wavefields,}
  {\protect\JournalTitle{Ultramicroscopy}} \textbf{157}, 57--64 (2015).

\bibitem{Ferri2010}
F.~Ferri, D.~Magatti, L.~A. Lugiato, and A.~Gatti, \enquote{{Differential ghost
  imaging},} {\protect\JournalTitle{Physical Review Letters}} \textbf{104},
  1--4 (2010).

\bibitem{Gibson2020}
G.~M. Gibson, S.~D. Johnson, and M.~J. Padgett, \enquote{{Single-pixel imaging
  12 years on: a review},} {\protect\JournalTitle{Optics Express}} \textbf{28},
  28190--28208 (2020).

\bibitem{Shapiro2008}
J.~H. Shapiro, \enquote{Computational ghost imaging,}
  {\protect\JournalTitle{Phys. Rev. A}} \textbf{78}, 061802 (2008).

\bibitem{Kallepalli2021_SciReps}
A.~Kallepalli, J.~Innes, and M.~J. Padgett, \enquote{Compressed sensing in the
  far-field of the spatial light modulator in high noise conditions,}
  {\protect\JournalTitle{Scientific Reports}} \textbf{11}, 1--8 (2021).

\bibitem{Karp1992}
R.~M. Karp, \enquote{On-line algorithms versus off-line algorithms: How much is
  it worth to know the future?} in \emph{Proceedings of the IFIP 12th World
  Computer Congress on Algorithms, Software, Architecture - Information
  Processing '92, Volume 1 - Volume I,}  (1992), p. 416–429.

\bibitem{Gatti2004}
A.~Gatti, E.~Brambilla, M.~Bache, and L.~A. Lugiato, \enquote{Ghost imaging
  with thermal light: Comparing entanglement and classical correlation,}
  {\protect\JournalTitle{Phys. Rev. Lett.}} \textbf{93}, 093602 (2004).

\bibitem{Sun2012}
B.~Sun, S.~S. Welsh, M.~P. Edgar, J.~H. Shapiro, and M.~J. Padgett,
  \enquote{Normalized ghost imaging,} {\protect\JournalTitle{Optics Express}}
  \textbf{20}, 16892 (2012).

\bibitem{Pratt1969}
W.~Pratt, J.~Kane, and H.~Andrews, \enquote{Hadamard transform image coding,}
  {\protect\JournalTitle{Proceedings of the IEEE}} \textbf{57}, 58--68 (1969).

\bibitem{Sloane1976}
N.~J.~A. Sloane and M.~Harwit, \enquote{Masks for hadamard transform optics,
  and weighing designs,} {\protect\JournalTitle{Applied Optics}} \textbf{15},
  107--114 (1976).

\bibitem{Zhang2017}
Z.~Zhang, X.~Wang, G.~Zheng, and J.~Zhong, \enquote{Hadamard single-pixel
  imaging versus fourier single-pixel imaging,} {\protect\JournalTitle{Optics
  Express}} \textbf{25}, 19619--19639 (2017).

\bibitem{Kallepalli2021Ghost}
A.~Kallepalli, D.~Stellinga, M.-J. Sun, R.~Bowman, E.~Rotunno, V.~Grillo, and
  M.~Padgett, \enquote{Ghost imaging with electron microscopy inspired,
  non-orthogonal phase masks,} {\protect\JournalTitle{{Research Square
  Preprint}}}  (2021).

\bibitem{Popa2004}
C.~Popa and R.~Zdunek, \enquote{Kaczmarz extended algorithm for tomographic
  image reconstruction from limited-data,} {\protect\JournalTitle{Mathematics
  and Computers in Simulation}} \textbf{65}, 579--598 (2004).

\bibitem{Bautu2006}
A.~Bautu, E.~Bautu, and C.~Popa, \enquote{A weighted kaczmarz algorithm in
  image reconstruction,} in \emph{Proceedings of the Fifth Workshop on
  Mathematical Modelling of Environmental and Life Sciences Problems,}  (2006),
  pp. 43--50.

\bibitem{Strohmer2008}
T.~Strohmer and R.~Vershynin, \enquote{A randomized kaczmarz algorithm with
  exponential convergence,} {\protect\JournalTitle{Journal of Fourier Analysis
  and Applications}} \textbf{15}, 262--278 (2008).

\bibitem{Pires2016}
R.~G. Pires, D.~R. Pereira, L.~A. Pereira, A.~F. Mansano, and J.~P. Papa,
  \enquote{Projections onto convex sets parameter estimation through harmony
  search and its application for image restoration,}
  {\protect\JournalTitle{Natural Computing}} \textbf{15}, 493--502 (2016).

\bibitem{Sun2020}
M.-L. Sun, C.-Q. Gu, and P.-F. Tang, \enquote{On randomized sampling kaczmarz
  method with application in compressed sensing,}
  {\protect\JournalTitle{Mathematical Problems in Engineering}} \textbf{2020},
  1--11 (2020).

\bibitem{Jesacher2007}
A.~Jesacher, A.~Schwaighofer, S.~F\"{u}rhapter, C.~Maurer, S.~Bernet, and
  M.~Ritsch-Marte, \enquote{Wavefront correction of spatial light modulators
  using an optical vortex image,} {\protect\JournalTitle{Optics Express}}
  \textbf{15}, 5801--5808 (2007).

\bibitem{Cizmar2010}
T.~Čižmár, M.~Mazilu, and K.~Dholakia, \enquote{In situ wavefront correction
  and its application to micromanipulation,} {\protect\JournalTitle{Nature
  Photonics}} \textbf{4}, 388--394 (2010).

\bibitem{Turtaev2017}
S.~Turtaev, I.~T. Leite, K.~J. Mitchell, M.~J. Padgett, D.~B. Phillips, and
  T.~Čižmár, \enquote{Comparison of nematic liquid-crystal and dmd based
  spatial light modulation in complex photonics,} {\protect\JournalTitle{Optics
  Express}} \textbf{25}, 29874--29884 (2017).

\end{thebibliography}

\end{document}